\documentclass[prx,noshowpacs,print,twocolumn]{revtex4}
\usepackage{amssymb}
\usepackage{amsmath}
\usepackage{graphicx}
\usepackage{dcolumn}
\usepackage{bm}
\usepackage{txfonts}
\usepackage{appendix}
\usepackage[dvips]{color}
\usepackage[colorlinks,linkcolor=red,citecolor=blue]{hyperref}

\begin{document}

\title{Anomalous quantum scattering and transport of electrons with
Mexican-hat dispersion induced by electrical potential}
\author{Jiating Yao$^{1}$}
\author{Benliang Zhou$^{1}$}
\author{Xiaoying Zhou$^{2,1}$}
\email{xyzhou@hnust.edu.cn}
\author{Xianbo Xiao$^{3}$}
\email{20101034@jxutcm.edu.cn}
\author{Guanghui Zhou$^{4,1}$}
\affiliation{$^1$Department of Physics, Key Laboratory for Low-Dimensional Structures and
Quantum Manipulation (Ministry of Education), Hunan Normal University,
Changsha 410081, China.}
\affiliation{$^2$School of Physics and Electronics, Hunan Provincial Key Laboratory of
Intelligent Sensors and Advanced Sensor Materials, Hunan University of
Science and Technology, Xiangtan 411201, China.}
\affiliation{$^3$School of Computer Science, Jiangxi University of Chinese Medicine,
Nanchang, 330004, China.}
\affiliation{$^4$Department of Physics, Shaoyang University, Shaoyang 422001, China.}

\begin{abstract}
We theoretically study the quantum scattering and transport of
electrons with Mexican-hat dispersion through both step and rectangular
potential barriers by using the transfer matrix method.
Owing to the torus-like iso-energy lines of the Mexican-hat dispersion, we observe the presence of double reflections and double transmissions in both two different barrier scenarios, i.e., the normal reflection (NR), retro-reflection (RR), normal transmission (NT), and specular transmission (ST).
For the step potential with electrons incident from the large wavevector, the transmission is primarily governed by NT with nearly negligible ST, while the reflection is dominant by RR (NR) within (outside) the critical angle. Additionally, for electrons incident from the small wavevector, the NT can be reduced to zero by adjusting the barrier, resulting in a significant enhancement of ST and RR.
For the rectangular barrier, the transmission and reflection spectra resemble those of the step barrier, but there are two kinds of resonant tunneling which can lead to perfect NT or ST. There exists a negative differential conductance (NDC) effect in the conductance spectrum. The conductance and the peak-to-valley ratio of the NDC effect can be effectively controlled by adjusting the height and width of the barrier as well as the incident energy.
Our results provide a deeper understanding of the electron states governed by the Mexican-hat dispersion.
\end{abstract}

\maketitle


\section{INTRODUCTION}

The discovery of graphene in 2004 marked the beginning of extensive research
on two-dimensional (2D) materials \cite{Novoselov1}. Since then, 2D
materials have rapidly gained prominence in condensed matter physics due to
their critical dimensions, atomic thickness, and diverse band structures
\cite{Sahin,Shaoliang Yu,Gibertini}. Many excellent and exotic properties
associated to 2D materials can not be found in their bulk counterparts \cite%
{Sahin,Shaoliang Yu,Gibertini,Gao,Hui Cai,Tuan}. To date, a considerable number of new 2D materials have been both theoretically predicted and experimentally synthesized \cite{Gao,Hui Cai,Tuan}. The band structures of 2D materials
change qualitatively as the thickness is reduced to few or monolayer. The
most famous example is graphene, which has a linear dispersion near the
Fermi level, requiring a description by 2D Dirac equation \cite{Novoselov}.
Indirect-direct band gap transition occurs when the thickness of 2D
transition metal dichalcogenides (TMDCs) decreases to monolayer \cite{Mak}.
For some 2D materials, the band structure undergos a transformation from a
parabolic dispersion into a Mexican-hat shape as their thickness decreases
to monolayer, which is also known as a Lifshiftz transition \cite{Zolyomi}.
Several 2D electron systems possess Mexican-hat dispersion in the low energy
regime, as confirmed by experiments \cite%
{Min,Jiang,Falko2016,Jappor,Ariapour,Demirci,Bandurin,Kibirev}. Notable
examples include the gated bilayer graphene \cite{Min}, the inverted
InAs/GaSb double quantum well \cite{Jiang}, and the ultrathin 2D group IIIA
metal chalcogenides MX (M=Ga, In and X=S, Se, Te) \cite%
{Falko2016,Jappor,Ariapour,Demirci,Bandurin,Kibirev}. Interestingly, the
Mexican-hat shaped bands can be tuned by strain or electric fields. For
instance, in monolayer InSe, the brim of the Mexican-hat bands can be
continuously manipulated by strain \cite{HuTing,Zhouma1} or a vertical
electric field \cite{Skinner,XiaoXB}, accompanying with an indirect-direct
band gap transition.

A prominent feature of Mexican-hat-shaped bands is the singularity in the
density of states at or near the band edge, where the density of modes
exhibits an abrupt step function \cite%
{Nurhuda,Seixas,Cao,Heremans,Ghobadi,Houssa,Xu-Jin Ge,Drummond,Wickramaratne}. The
distorted density of states results in exotic electric and magnetic
properties in electron systems with sombrero-shaped dispersion \cite%
{Nurhuda,Seixas,Cao,Heremans,Ghobadi,Houssa,Xu-Jin Ge,Drummond,Wickramaratne}, such
as the multi-ferroic effect in $\alpha$-SnO \cite{Seixas}, tunable magnetism
and half-metallicity monolayer GaSe \cite{Cao}, the enhanced thermoelectric
efficiency in few-layer III-VI materials \cite{Heremans,Houssa,Xu-Jin
Ge,Wickramaratne}, and the Lifshitz transitions in 2D InSe \cite{Drummond}.
In addition, the interplay between the Mexican-hat dispersion and the
Rashba/Dresselhaus spin-orbit coupling may result in the spin-charge
conversion \cite{Zhouma2} and spin polarization \cite{Zhouma3}.
In conventional semiconductors, the Fermi surface is a circle, and electron
scattering against potential barrier follows a Snell's law akin to optics
\cite{Griffiths}, resulting in specular reflection and normal transmission.
This behavior is also observed in graphene. However, graphene p-n junctions
exhibit exotic negative refraction \cite{Cheianov,GHLee}, which is useful to focus the
electron beam. For electrons with Mexican-hat shaped bands, the Fermi
surfaces are two concentric circles, similar to a torus. There are two extra
scattering channels, which will give rise to novel scattering features when
electrons encounter a potential barrier. However, this topic remains unexplored, and the underlying scattering mechanisms are exclusive.

In this work, we theoretically study the quantum scattering and transport of
electrons with Mexican-hat dispersion through both step and rectangular
potential barriers by using the transfer matrix method.
Owing to the torus-like iso-energy lines of the Mexican-hat dispersion, we observe the presence of double reflections and double transmissions in both two different barrier scenarios, i.e., the normal reflection (NR), retro-reflection (RR), normal transmission (NT), and specular transmission (ST).
For the step potential with electrons incident from the large wavevector, the transmission is primarily governed by NT with nearly negligible ST, while the reflection is dominant by RR (NR) within (outside) the critical angle. Additionally, for electrons incident from the small wavevector, the NT can be reduced to zero by adjusting the barrier, resulting in a significant enhancement of ST and RR.
For the rectangular barrier, the transmission and reflection spectra resemble those of the step barrier, but there are two kinds of resonant tunneling which can lead to perfect NT or ST. There exists a negative differential conductance (NDC) effect in the conductance spectrum. The conductance and the peak-to-valley ratio of the NDC effect can be effectively controlled by adjusting the height and width of the barrier as well as the incident energy.

The rest of this paper is organized as follows. In Sec. II, the quartic
model and transfer matrix method for the Mexican-hat-dispersion electrons
are introduced. In Sec. III, numerical results and discussions for the
scattering and transport of Mexican-hat-dispersion electrons are presented.
Finally, a brief summary of the results is given in Sec. IV.

\section{MODEL AND METHODS}

The simplest Mexican-hat dispersion takes the form of the quartic model \cite{MaZhou1}%
\begin{equation}
E\left( \bm{\emph{k}}\right) =h_{1}\bm{\emph{k}}^{2}+h_{2}\bm{\emph{k}}%
^{4}-E_{T},  \label{Eh}
\end{equation}%
where $h_{1}$ and $h_{2}$ are constants meeting the conditions $h_{1}<0$ and
$h_{2}>0$ for the conduction band while $h_{1}>0$ and $h_{2}<0$ for the
valence band. The energy $E_{T}$=$-h_{1}^{2}/4h_{2}$ represents the
height of the Mexican hat, and $\textbf{k}$=$\left( k_{x},k_{y}\right) $ is
the wave vector. The single-particle effective Hamiltonian corresponding to
Eq. \ref{Eh} can be written as
\begin{equation}
H\left( \bm{\emph{k}}\right) =-h_{1}\left( \frac{\partial ^{2}}{%
\partial x^{2}}+\frac{\partial ^{2}}{\partial y^{2}}\right) +h_{2}\left(
\frac{\partial ^{2}}{\partial x^{2}}+\frac{\partial ^{2}}{\partial y^{2}}%
\right) ^{2}-E_{T},  \label{H}
\end{equation}%
which is obtained by replacing the wave vectors ($k_{x}$, $k_{y}$) with
differential operators ($-i\partial _{x}$, $-i\partial _{y}$) in Eq. (\ref%
{Eh}). For a fixed electron energy $E$, we have
\begin{equation}
k^{2}=\frac{-h_{1}\pm \sqrt{h_{1}^{2}+4h_{2}\left( E+E_{T}\right) }}{2h_{2}},
\end{equation}%
which means the iso-energy lines, i.e., the Fermi surfaces, are two
concentric circles. Their radii are $k_{\pm }$=$\left[ \left( -h_{1}\pm
\gamma \right) /2h_{2}\right] ^{1/2}$with $\gamma$ =$\sqrt{h_{1}^{2}+4h_{2}%
\left( E+E_{T}\right) }$, where $k_{+}\left( k_{-}\right) $ indicates the
small (big) iso-energy circles. The probability current density along the
two direction can be obtained by $\mathbf{j}=-\frac{i}{\hbar }\left[ \mathbf{%
r},H\right] $, which are given by
\begin{equation}
j_{x/y}=k_{x/y}\left( h_{1}+2h_{2}k^{2}\right) =\pm \gamma k_{x/y},
\end{equation}%
Therefore, the direction of the probability current density $\mathbf{j}$ is
identical (opposite) to the direction of the wave vector $\mathbf{k}$ on the
small (big) iso-energy circle.
\begin{figure}[tbp]
\includegraphics[width=0.5\textwidth]{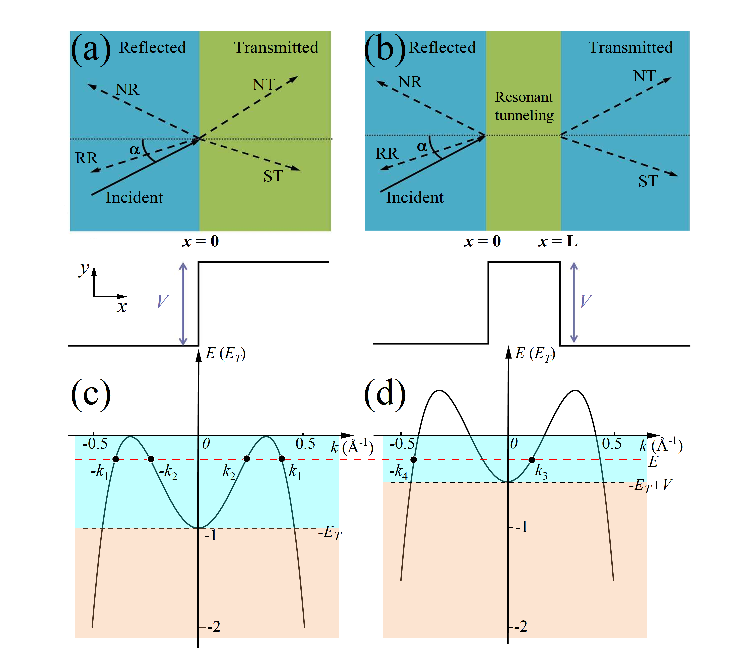}
\caption{(Color online) Schematic illustration of the scattering processes of the
Mexican-hat-dispersion electrons at the interfaces of a step
(a) and rectangular (b) potential barrier. The solid and
the dashed arrows denote the group velocities of the incident and the
outgoing electrons, respectively. The height of the potential barrier is $V$
and the barrier width in (b) is $L$. Energy bands of the scattering regions
without (c) and with (d) potential barrier.}
\label{fig1}
\end{figure}

\subsection{Scattering at the interface of a step potential barrier}

Firstly, we consider the scattering at interface of a step potential barrier
along the $x$-direction with the height $V$ as shown in Fig. \ref{fig1}(a).
According to the Mexican-hat dispersion shown in Fig. \ref{fig1}(c), we can
divide the incident electron energy $E$ into two distinct regimes, each
giving rise to different scattering mechanisms. The first regime is $%
-E_{T}<E<0$, as indicated by the light cyan region in Fig. \ref{fig1}(c), in
which the electron energy is confined within the Mexican-hat. In this
scenario, at a specific energy level, there exist four modes as shown in
Fig. \ref{fig1}(c), two propagating forward ($-k_{1}$ and $k_{2}$
states) and two propagating backward ($-k_{2}$ and $k_{1}$ states).
Consequently, one will observe retro-reflection and specular transmission,
i.e., the negative refraction, in addition to the normal reflection and
transmission, which is absent in the electron scattering of conventional 2D
electron gas. Since there are two modes propagating forward along the $x$%
-direction, we should discuss them separately. If the electron is incident
with the wave vector $-k_{1}$, the wave functions in the incident/reflection
($x<0$) and the transmission ($x>0$) regions in Fig. \ref{fig1}(a) can be
expressed respectively as
\begin{equation}
\begin{aligned} &
\psi_{1}(x)=e^{-ik_{x1}x}+r_{1}e^{ik_{x1}x}+r_{2}e^{-ik_{x2}x},\\ &\psi
_{2}(x)=t_{1}e^{ik_{x3}x}+t_{2}e^{-ik_{x4}x}, \end{aligned}  \label{WF1}
\end{equation}%
where $k_{xi}$=$(k_i^2-k_y^2)^{1/2}, i=1,2,3,4$. The wavevectors $k_i$ are $k_1$=$k_{+}$, $k_2$=$k_{-}$,
$k_{3}=[(-h_{1}+\chi)/2h_{2}]^{1/2}$, and $k_{4}=[(-h_{1}-\chi)/2h_{2}]^{1/2}$
with $\chi =[h_{1}^{2}+4h_{2}\left( E_{T}+E-V\right)]^{1/2}$. We have omitted the common
factor $e^{ik_{y}y}$ in Eq. (\ref{WF1}) because the wavevector $k_{y}$ is
conserved arising from the translation invariance along the $y$-direction.

The type of each scattering ($r_{i}$ and $t_{i},i=1,2$) can be classified
according to the relative orientation between the wave vector and the group
velocity of both reflected and transmitted electrons. The group velocity for
a given $k$ states is $\bm{\upsilon}_{g}\left( k\right) =\frac{1}{\hbar }%
\triangledown _{\bm{\emph{k}}}E_{h}\left( \bm{\emph{k}}\right) $. We can use
the product of the wave vector and the group velocity to manifest the
relative orientation between them. For the reflected electron state with
wave vector $\bm{\emph{k}}_{x1}$, we have $\bm{\emph{k}}_{x1}\cdot %
\bm{\upsilon}_{g}\left( k_{1}\right) <0$, which is consistent with that of
the incident electron state. Therefore, this is a normal reflection (NR)
with magnitude $r_{1}$ , as indicated in Fig. \ref{fig1}(a). However, for
another reflected electron state wave vector $-\bm{\emph{k}}_{x2}$, we have $%
-\bm{\emph{k}}_{x2}\cdot \bm{\upsilon}_{g}\left( -k_{2}\right) >0$, which is
opposite to that of the incident electron state so that it is a
retro-reflection (RR) with magnitude $r_{2}$. Similarly, for the transmitted
electron states $k_{3}$ and $-k_{4}$, we have $\bm{\emph{k}}_{x3}\cdot %
\bm{\upsilon}_{g}\left( k_{3}\right) >0$ and $-\bm{\emph{k}}_{x4}\cdot %
\bm{\upsilon}_{g}\left( -k_{4}\right) <0$, which corresponding to the
specular transmission (ST) and normal transmission (NT) with magnitude $%
t_{1} $ and $t_{2}$, respectively.

Since the Hamiltonian in Eq. (\ref{Eh}) contains the quartic term of $k$,
the wave functions in both regions and their first to third order
derivatives should be continuous at the scattering interface $x=0$ \cite{Ruhl},
i.e., $\psi _{1}\left( 0\right) =\psi _{2}\left( 0\right) ,$ $\psi
_{1}^{\prime }\left( 0\right) =\psi _{2}^{\prime }\left( 0\right) ,$ $\psi
_{1}^{\prime \prime }\left( 0\right) =\psi _{2}^{\prime \prime }\left(
0\right) ,$ $\psi _{1}^{\prime \prime \prime }\left( 0\right) =\psi
_{2}^{\prime \prime \prime }\left( 0\right) $. Once these boundary
conditions are considered, we obtain the reflection and transmission
coefficients as follows
\begin{equation}
r_{1}=\frac{\left( k_{x1}-k_{x2}\right) \left( k_{x1}-k_{x4}\right) \left(
k_{x1}+k_{x3}\right) }{\left( k_{x1}-k_{x3}\right) \left(
k_{x1}+k_{x2}\right) \left( k_{x1}+k_{x4}\right) },
\end{equation}%
\begin{equation}
r_{2}=-2k_{x1}\frac{\left( k_{x1}-k_{x4}\right) \left( k_{x1}+k_{x3}\right)
}{\left( k_{x2}-k_{x4}\right) \left( k_{x1}+k_{x2}\right) \left(
k_{x2}+k_{x3}\right) },
\end{equation}%
\begin{equation}
t_{1}=2k_{x1}\frac{\left( k_{x1}-k_{x2}\right) \left( k_{x1}-k_{x4}\right) }{%
\left( k_{x1}-k_{x3}\right) \left( k_{x2}+k_{x3}\right) \left(
k_{x3}+k_{x4}\right) },
\end{equation}%
\begin{equation}
t_{2}=-2k_{x1}\frac{\left( k_{x1}-k_{x2}\right) \left( k_{x1}+k_{x3}\right)
}{\left( k_{x2}-k_{x4}\right) \left( k_{x1}+k_{x4}\right) \left(
k_{x3}+k_{x4}\right) }.
\end{equation}

Further, the probability current density operator $j$ in the Mexican hat
dispersion can be derived by $j=-\frac{i}{\hbar }\left[r,H\right] $ so that
the $x$-component of the probability current density is given as $j_{x}=k_{x}%
\left[ h_{1}+2h_{2}\left(k_{x}^{2}+k_{y}^{2}\right) \right]$. According to
the conservation of the probability current, the reflection probabilities of
the NR and RR are obtained as
\begin{equation}
\begin{aligned} &R_{1}=\left\vert r_{1}\right\vert ^{2},\\
&R_{2}=\operatorname{Re}\left[ \frac{k_{x2}}{k_{x1}}\right] \left\vert
r_{2}\right\vert ^{2}\left\vert \frac{h_{1}+2h_{2}\left(
k_{x2}^{2}+k_{y}^{2}\right) }{h_{1}+2h_{2}\left( k_{x1}^{2}+k_{y}^{2}\right)
}\right\vert. \end{aligned}
\end{equation}%
The transmission probabilities of the ST and NT are
\begin{equation}
\begin{aligned} &T_{1}=\operatorname{Re}\left[ \frac{k_{x3}}{k_{x1}}\right]
\left\vert t_{1}\right\vert ^{2}\left\vert \frac{h_{1}+2h_{2}\left(
k_{x3}^{2}+k_{y}^{2}\right) }{h_{1}+2h_{2}\left( k_{x1}^{2}+k_{y}^{2}\right)
}\right\vert,\\ &T_{2}=\operatorname{Re}\left[ \frac{k_{x4}}{k_{x1}}\right]
\left\vert t_{2}\right\vert ^{2}\left\vert \frac{h_{1}+2h_{2}\left(
k_{x4}^{2}+k_{y}^{2}\right) }{h_{1}+2h_{2}\left( k_{x1}^{2}+k_{y}^{2}\right)
}\right\vert,\end{aligned}
\end{equation}
respectively.

Similarly, if the electron is incident with another wave vector $k_{2}$, as
indicated in Fig. \ref{fig1}(c). For electron energies within the
Mexican-hat regime. In this case, the wave functions in the two regions can
be expressed respectively as
\begin{equation}
\begin{aligned} &\psi
_{1}=e^{ik_{x2}x}+\widetilde{r_{1}}e^{-ik_{x2}x}+%
\widetilde{r_{2}}e^{ik_{x1}x},\\ &\psi
_{2}=\widetilde{t_{1}}e^{ik_{x3}x}+\widetilde{t_{2}}e^{-ik_{x4}x}.%
\end{aligned}
\end{equation}%
After a similar calculation procedure, the reflection and the transmission
coefficients $\widetilde{r_{1}},\widetilde{r_{2}},\widetilde{t_{1}},%
\widetilde{t_{2}}$ are obtained as
\begin{equation}
\begin{aligned} &\widetilde{r_{1}} =\frac{-\left( k_{x1}-k_{x2}\right)
\left( k_{x2}-k_{x3}\right) \left( k_{x2}+k_{x4}\right) }{\left(
k_{x2}-k_{x4}\right) \left( k_{x1}+k_{x2}\right) \left( k_{x2}+k_{x3}\right)
}, \\ &\widetilde{r_{2}} =-2k_{x2}\frac{\left( k_{x2}-k_{x3}\right) \left(
k_{x2}+k_{x4}\right) }{\left( k_{x1}-k_{x3}\right) \left(
k_{x1}+k_{x2}\right) \left( k_{x1}+k_{x4}\right) }, \\&\widetilde{t_{1}}
=2k_{x2}\frac{\left( k_{x1}-k_{x2}\right) \left( k_{x2}+k_{x4}\right)
}{\left( k_{x1}-k_{x3}\right) \left( k_{x2}+k_{x3}\right) \left(
k_{x3}+k_{x4}\right) }, \\ &\widetilde{t_{2}} =-2k_{x2}\frac{\left(
k_{x1}-k_{x2}\right) \left( k_{x2}-k_{x3}\right) }{\left(
k_{x2}-k_{x4}\right) \left( k_{x1}+k_{x4}\right) \left( k_{x3}+k_{x4}\right)
}. \end{aligned}
\end{equation}%
The corresponding probabilities for the NR, RR, ST and NT are
\begin{equation}
\begin{aligned} &R_{1} =\left\vert \widetilde{r_{1}}\right\vert ^{2} ,\\
&R_{2} =\operatorname{Re}\left[ \frac{k_{x1}}{k_{x2}}\right] \left\vert
\widetilde{r_{2}}\right\vert ^{2}\left\vert \frac{h_{1}+2h_{2}\left(
k_{x1}^{2}+k_{y}^{2}\right) }{h_{1}+2h_{2}\left( k_{x2}^{2}+k_{y}^{2}\right)
}\right\vert, \\ &T_{1} =\operatorname{Re}\left[
\frac{k_{x4}}{k_{x2}}\right] \left\vert \widetilde{t_{2}}\right\vert
^{2}\left\vert \frac{h_{1}+2h_{2}\left( k_{x4}^{2}+k_{y}^{2}\right)
}{h_{1}+2h_{2}\left( k_{x2}^{2}+k_{y}^{2}\right) }\right\vert, \\ &T_{2}
=\operatorname{Re}\left[ \frac{k_{x3}}{k_{x2}}\right] \left\vert
\widetilde{t_{1}}\right\vert ^{2}\left\vert \frac{h_{1}+2h_{2}\left(
k_{x3}^{2}+k_{y}^{2}\right) }{h_{1}+2h_{2}\left( k_{x2}^{2}+k_{y}^{2}\right)
}\right\vert, \end{aligned}
\end{equation}
respectively.

According to the transmission probabilities obtained above, the total
conductance of the junction at zero temperature is
\begin{equation}
G=\sum\limits_{s} 2G_{0}\int_{-\frac{\pi }{2}}^{\frac{\pi }{2}}\left(
T_{1}^{s}+T_{2}^{s}\right) \cos \alpha d\alpha,
\end{equation}
where $s=-k_{x1},~k_{x2}$ denotes different incident wave vectors, $%
\alpha=\arctan(k_{y}/k_{x})$ is the incident angle, and $G_{0}=\frac{e^{2}}{h%
}N_{0}\left(E\right)$. Here $N_{0} \left(E\right) =\frac{W\left\vert
k_{s}\right\vert }{\pi }$ is the transverse modes in a sheet of Mexican hat
dispersion materials with width $W$.

For electron energy outside the Mexican-hat regime, $E<-E_{T}$, as indicated
by the light yellow region in Fig. \ref{fig1}(c), the scattering process
returns to the normal case. There is only one incident, transmission and
reflection state, respectively and the wave functions in the two regions are
reduced to
\begin{equation}
\begin{aligned} &\psi _{1}=e^{-ik_{x1}x}+r_{1}e^{ik_{x1}x},\\ &\psi
_{2}=t_{2}e^{-ik_{x4}x}.\end{aligned}
\end{equation}
Considering the continuity of the wave function at the interface $%
1+r_{1}=t_{2}$, and the conservation of the probability current requires
that
\begin{equation*}
\begin{aligned} &-k_{x1}\left( h_{1}+2h_{2}\left(
k_{x1}^{2}+k_{y}^{2}\right) \right) +k_{x1}\left( h_{1}+2h_{2}\left(
k_{x1}^{2}+k_{y}^{2}\right)\right)R_{1}\\ &=-k_{x4}\left( h_{1}+2h_{2}\left(
k_{x4}^{2}+k_{y}^{2}\right) \right) T_{2}, \end{aligned}
\end{equation*}
we obtain%
\begin{equation}
r_{1} =\frac{k_{x1}\left( h_{1}+2h_{2}\left( k_{x1}^{2}+k_{y}^{2}\right)
\right) -k_{x4}\left( h_{1}+2h_{2}\left( k_{x4}^{2}+k_{y}^{2}\right) \right)
}{k_{x4}\left( h_{1}+2h_{2}\left( k_{x4}^{2}+k_{y}^{2}\right) \right)
+k_{x1}\left( h_{1}+2h_{2}\left( k_{x1}^{2}+k_{y}^{2}\right) \right) },
\end{equation}
\begin{equation}
t_{2} =\frac{2k_{x4}\left( h_{1}+2h_{2}\left( k_{x4}^{2}+k_{y}^{2}\right)
\right) }{k_{x4}\left( h_{1}+2h_{2}\left( k_{x4}^{2}+k_{y}^{2}\right)
\right) +k_{x1}\left( h_{1}+2h_{2}\left( k_{x1}^{2}+k_{y}^{2}\right) \right)
},
\end{equation}%
and
\begin{equation}
\begin{aligned} &R_{1} =\left\vert r_{1}\right\vert ^{2}, \\
&T_{2}=\operatorname{Re}\left[ \frac{k_{x4}}{k_{x1}}\right] \left\vert
\frac{h_{1}+2h_{2}\left( k_{x4}^{2}+k_{y}^{2}\right) }{h_{1}+2h_{2}\left(
k_{x1}^{2}+k_{y}^{2}\right) }\right\vert \left\vert t_{2}\right\vert ^{2}.
\end{aligned}
\end{equation}

\subsection{Scattering at the interfaces of a rectangular potential barrier}

Secondly, we consider the scattering at the interfaces of a rectangular
potential barrier $V\left(x\right)=V[\Theta\left(x\right)-\Theta\left(x-L%
\right)]$, as shown in Fig. \ref{fig1} (b), where $\Theta\left(x\right)$ is
a step function. In this case, the whole space is separated into three
regions, i.e. the incident/reflection ($x<0$), the resonant tunneling ($%
0<x<L $), and the transmission ($x>L$) regions. For the injected electron
with energy $E$ and wave vector $k_{x2}$, the wave functions in the three
regions are given as
\begin{equation}
\begin{aligned}
&\psi_{1}=e^{ik_{x2}x}+{r_{1}}e^{-ik_{x2}x}+{r_{2}}e^{ik_{x1}x},\\
&\psi_{2}=A_{1}e^{ik_{x3}x}+B_{1}e^{-ik_{x4}x}+A_{2}e^{ik_{x4}x}+B_{2}e^{-ik_{x3}x},\\ &\psi_{3}=t_{1}e^{ik_{x2}x}+t_{2}e^{-ik_{x1}x}. \end{aligned}
\end{equation}
The reflection and transmission coefficients $r_{1}$, $r_{2}$, $t_{1}$, $%
t_{2}$ and the transmission probabilities $R_{1}$, $R_{2}$, $T_{1}$, $T_{2}$
can also be obtained by using the similar calculations in above subsection.
However, complex algebraic operations are required to obtain the analytical
forms of the reflection and transmission probabilities. Further, the
analytical results are also complex and tedious. Therefore, the detail
analytical calculation procedure and its results are not presented for
conciseness while the numerical results are given in the following section.
\begin{figure*}[tbp]
\centering
\includegraphics[width=0.98\textwidth]{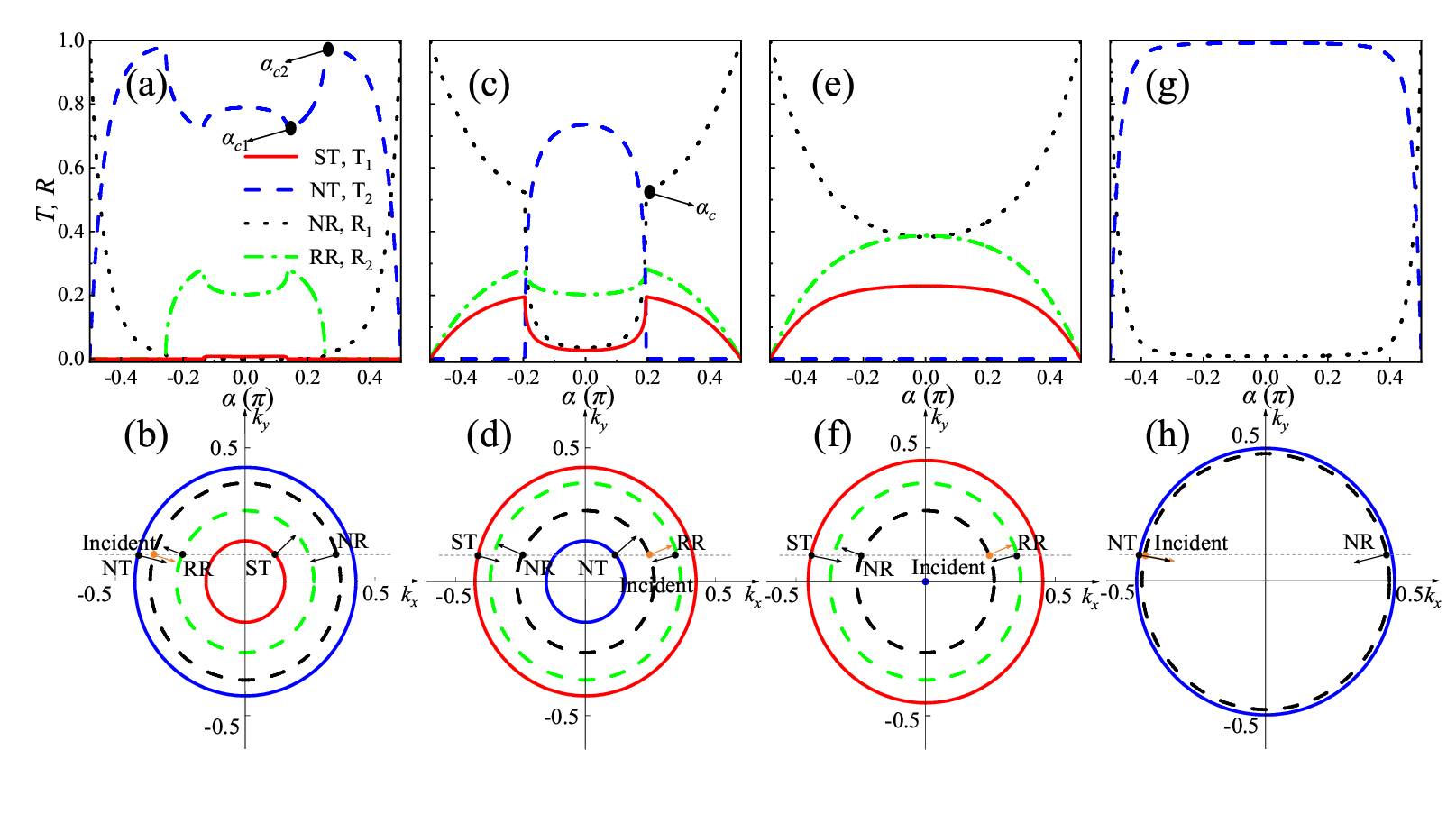}
\caption{(Color online) Reflection and the transmission spectra as a
function of the incident angle $\alpha$ for the scattering induced by a step
potential barrier. The incident energy, barrier height, and incident wave vector $k_i$ are set as (a) $E$=$-0.1$, $V$=$0.5$, $k_i$= $-k_{1}$, (c) $E$=$-0.1$, $V$=$0.5$, $k_i$= $k_{2}$, (e) $E$=$-0.1$, $V$=$0.9$, $k_i$= $k_{2}$, and (g) $E$=$-1.5$, $V$=$0.5$, $k_i$= $-k_{1}$. The corresponding iso-energy circles for the incident and transmitted regions are plotted in (b), (d), (f) and (h).
}
\label{fig2}
\end{figure*}

\section{Numerical results and discussions}

In what follows we show some numerical examples for the electron scattering
and transport properties of electrons with Mexican-hat dispersion. Although,
the physics discussed here do not depend on the band parameters of the
Mexican-hat-dispersion, to provide specificity, the parameters of the energy
band in monolayer InSe are chosen as an illustrative example, namely $%
h_{1}=1.478$ eV $\mathring{\mathrm{A}}^{2}$, $h_{2}=-7.219$ eV $\mathring{%
\mathrm{A}}^{4}$, and $E_{T}=75.7$ meV \cite{Ma Zhou1}. The incident energy of the electron $E$ and the hight of the electric potential $V$ are in unit of $E_T$, and the barrier width of the rectangular potential is in unit of nm.

In Fig. \ref{fig2}(a), the reflection and transmission probabilities are
plotted as a function of the incident angle at the interface of a step
potential barrier. The electrons are incident with wave vector $-k_1$. The barrier height and incident energy are set as $V$=0.5 and $E$=$-$0.1, respectively. The corresponding iso-energy
lines for the incident and transmitted regions are presented in Fig. \ref%
{fig2}(b). The radii of these concentric circles from the innermost to the
outermost are $k_{3}=\left[ \left( -h_{1}+\chi \right) /2h_{2}%
\right] ^{1/2}$, $k_{2}=\left[ \left( -h_{1}+\gamma \right) /2h_{2}\right]
^{1/2}$, $k_{1}=\left[ \left( -h_{1}-\gamma \right) /2h_{2}\right] ^{1/2}$,
and $k_{4}=\left[ \left( -h_{1}-\chi \right) /2h_{2}\right] ^{1/2}$. Given
that both the reflection and transmission probabilities are even functions
concerning the incident angle $\alpha $, our discussions focus on the
results for $\alpha >0$. When $\alpha $ is small, as indicated by the
horizontal dashed line in Fig. \ref{fig2}(b), double reflections and double
transmissions can be generated because the four states are all
propagating modes. Among the four propagating modes indicated in Fig. \ref%
{fig2}(b), the incident state lies between the NT state and the RR state,
which indicates that the incident state is easier scattering into them as expressed in Eqs. (6)-(9).
Hence, the probabilities of NT ($T_{2}$) and RR ($R_{2}$) are much larger
than those of ST ($T_{1}$) and NR ($R_{1}$). Owing to the presence of ST, $%
T_{2}$ fails to achieve its maximum value at normal incidence, which is
different from the potential scattering in conventional semiconductors \cite%
{Griffiths}. $T_{1}$ increases with the escalating $\alpha $, whereas, $T_{2}
$ exhibits a gradual decreases as $\alpha $ increases until it reaches
the first critical angle $\alpha _{c1}$=$\arcsin (k_{3}/k_{1})$=0.14$\pi$ where
the horizontal dashed line becomes tangent to the red solid circle in Fig. %
\ref{fig2}(b). For $\alpha $$>$$\alpha _{c1}$, the ST state becomes
evanescent wave, then $T_{1}$ vanishes. However, $T_{2}$ increases to a peak
($\sim 1.0$) as $\alpha $ increases until it reaches the second critical
angle $\alpha _{c2}$=$\arcsin (k_{2}/k_{1})$=0.26$\pi $ where the horizontal
dashed line becomes tangent to the green dashed circle in Fig. \ref{fig2}%
(b). Finally, $T_{2}$ rapidly diminishes to zero as $\alpha $ approaches $%
\pi /2$. Meanwhile, $R_{2}$ displays opposite trends compared to $T_{2}$
with an increasing $\alpha $ up to $\alpha _{c2}$, shown by the green dash-
dotted line in Fig. \ref{fig2}(a). For $\alpha $$>$$\alpha _{c2}$, the RR
state becomes evanescent wave, and $R_{2}$ vanishes, leading to a sharply
increased $R_{1}$.

Figures \ref{fig2}(c) and \ref{fig2}(e) display the reflection and the transmission as
a function of the incident angle $\alpha$, in which electrons are incident with wave vector $k_2$. The incident energies are $E$=$-0.1$, and the barrier hight $V$ are $0.5$ and $0.9$ for Figs. \ref{fig2}(c) and \ref{fig2}(e), respectively. The corresponding iso-energy
lines for the incident and transmitted regions are presented in Figs. \ref{fig2}(d) and \ref{fig2}(f). The radii of these concentric are also in the order of $k_{3}$, $k_{2}$, $k_{1}$, and $k_{4}$.
For small barrier $V$=$0.5$, there are also four propagating states for small $\alpha$ as indicated by the horizontal dashed line in Fig. \ref{fig2}(d). Consequently, double reflections and double
transmissions also appear. Among the four propagating modes, the incident state lies between the NT state and the RR state,
which means that the incident state is easier scattering into them as expressed in Eq. (13).
Therefore, $T_{2}$ and $R_{2}$ are much larger
than $T_{1}$ and $R_{1}$.
Different from the results in Fig. \ref{fig2}(a), $T_{2}$ reaches its maximum value at normal incidence
and then decreases to zero when $\alpha$ increases to the critical angle $\alpha_c=\arcsin(k_{3}/k_{2})=0.195\pi$ where
the horizontal dashed line becomes tangent to the red solid circle in Fig. \ref{fig2}(d). For $\alpha$$>$$\alpha _{c}$, $T_{2}$ vanishes because the NT state becomes
evanescent wave. Whereas, $T_{1}$ increases with escalating $\alpha$ and achieve its maximum at  $\alpha_c$, then it gradually diminishes to zero as $\alpha $ approaches $\pi/2$.
Meanwhile, $R_{1}$ displays opposite trends compared to $T_{2}$, and $R_{2}$ shows the same trends compared to $T_{1}$
with the increasing of $\alpha$.
For large barrier $V$$\geq$$E_T+E$, such as $V$=$0.9$, the iso-energy circle corresponding to NT reduces to a point or disappears, and the ST state becomes evanescent wave. Hence, there are double reflections but only specular transmission in this case. Owing to the disappear of $T_{1}$, $T_{2}$ is greatly enhanced compared with the results in Figs. 2(a) and 2(c).
For electrons with incident energy satisfying $E$$<$$-E_T$,  there are only two modes of propagation, one incident state and one reflected state. The scattering caused by the step barrier is similar to that in conventional semiconductors. Figure \ref{fig2}(g) plots the reflection and
transmission spectra as a function of $\alpha$ with $E=-1.5$ and barrier hight $V=0.5$. In this case, there are only SR ($R_{1}$) and NT ($T_{2}$), as shown by the
black solid and the blue dotted lines, respectively. The $%
T_{2}$ decreases slowly first and then drop fast with the increase of the
$\alpha$, whereas, $R_{1}$ demonstrates a reversal behaviour, which is similar to the results in conventional semiconductors \cite{Griffiths}.
The reason is that only the NT and SR states are propagating modes
within the whole incident angle range, as shown by the iso-energy lines in Fig. \ref{fig2}(g).
\begin{figure}[tbp]
\centering
\includegraphics[width=0.48\textwidth]{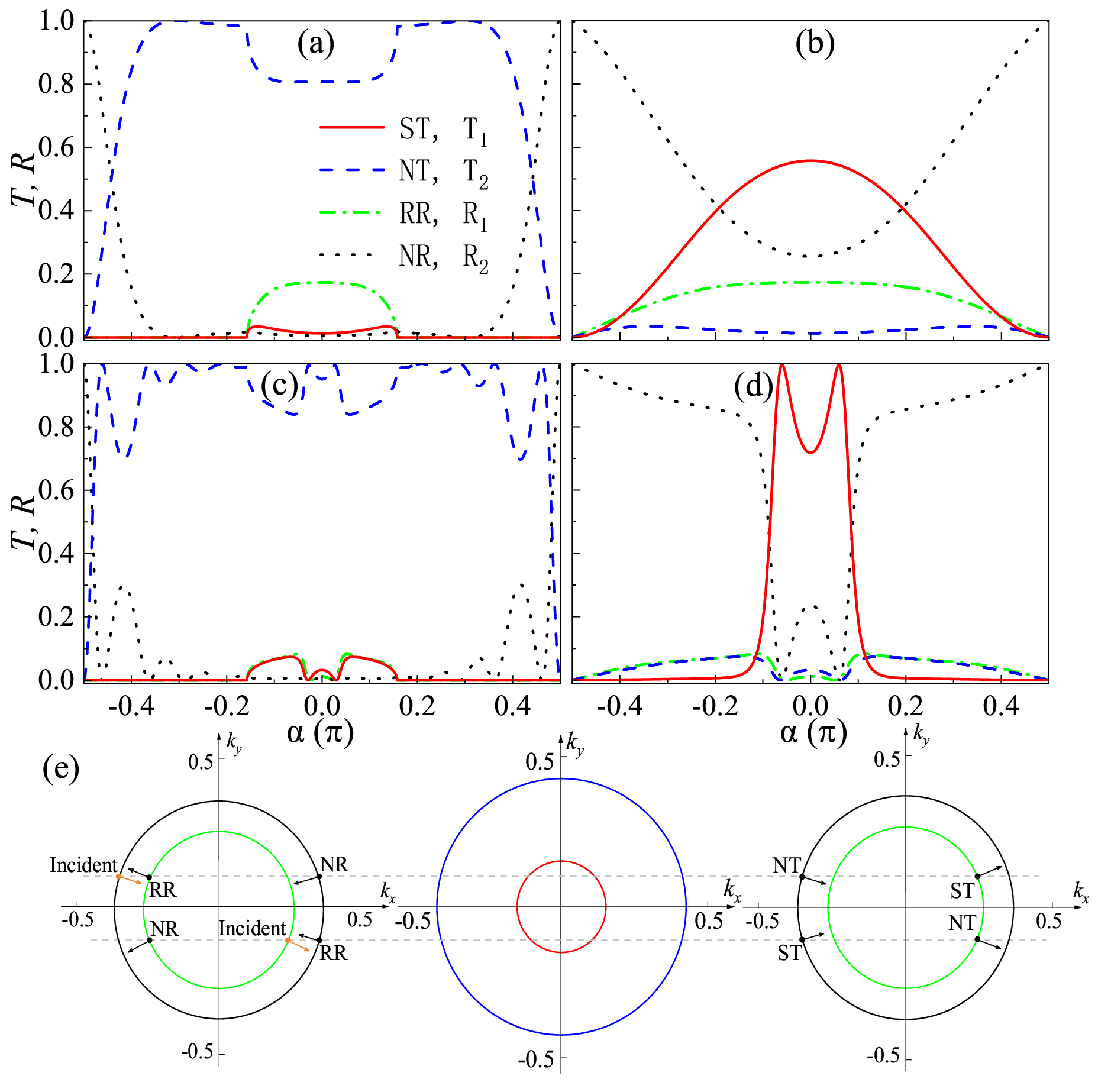}
\caption{(Color online)
(a)-(d) Reflection and the transmission spectra as a
function of the incident angle $\alpha$ for the scattering induced by a rectangle
potential barrier. The incident energy and the barrier height are -0.4 and 0.5, respectively. The barrier  width, and incident wave vector $k_i$ are set as (a) $L$=10 nm, $k_i$= $-k_{1}$, (b) $L$=10 nm, $k_i$= $k_{2}$, (c) $L$=50 nm, $k_i$= $-k_{1}$, and (d) $L$=50 nm, $k_i$=$k_{2}$. The corresponding iso-energy circles for the incident, center and transmitted regions are plotted in (e).
}
\label{fig3}
\end{figure}
\begin{figure}[t]
\centering
\includegraphics[bb=60 12 746 549, width=0.48\textwidth]{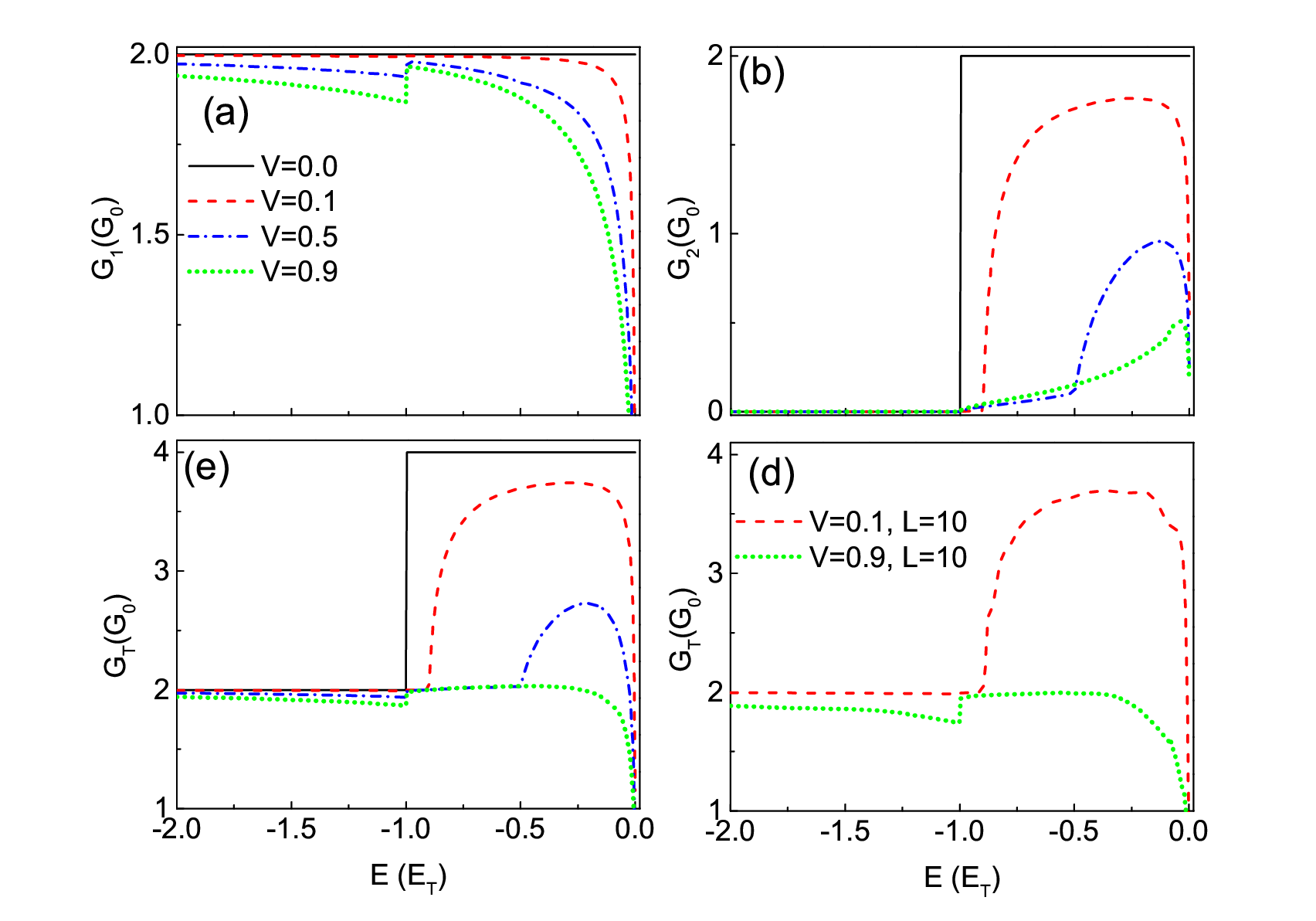}
\caption{(Color online) (a)-(c) The ballistic conductance as a function of incident energy for a step barrier with various potentials. (a) $G_1$ stands for electrons incident with wavevector $-k_1$. (b) $G_2$ denotes electrons incident with wavevector $k_2$. (c) $G_T$=$G_1$+$G_2$ is the total conductance. (d) The total conductance as a function of incident energy for a rectangle barrier with various potentials.}
\label{fig4}
\end{figure}

Next, let's turn to the scattering induced by a rectangle potential barrier. Figures \ref{fig3}(a)-(d) show the reflection and the transmission spectra as a
function of the incident angle $\alpha$ for the scattering induced by a rectangle
potential barrier for different barrier widths.
In comparison to the results of the stepped barrier, there are two notable features in the case of the rectangle barrier. One is that the probability of ST will increase because there are two forward propagating waves in the barrier region [see Eq. (20)], and both of them  produce ST at the interface $x=L$. Another is that there is resonant tunnelling. We will focus on these two features in the following discussions.

For electrons incident with wavevector $-k_1$, as shown in Figs. \ref{fig3}(a) and \ref{fig3}(c), there is a critical angle $\alpha_c$=$\arcsin(k_{3}/k_{1})$ where
the horizontal dashed line becomes tangent to the red solid circle in Fig. \ref{fig3}(e) in the reflection and transmission spectra. There are double reflections and transmissions for $0<\alpha<\alpha_c$ with enhanced  $T_1$ and reduced $R_2$ because both the two forward propagating waves produce ST at the interface $x=L$. However, $T_1$ and $R_2$ disappear for $\alpha$$>$$\alpha_c$, and $T_2$ ($R_1$) shows some resonate  peaks (dips), i.e., $T_2$=1 ($R_1$=0), as the increasing of $\alpha$ arising from the resonate tunnelling. The resonate condition can be obtained numerically and need to be discussed in two cases because it sensitively depends on $\alpha$.
For $0<\alpha<\alpha_c$, the wavevectors $k_{x3}$ and $k_{x4}$ both are real. We have numerical checked the resonate condition and found it requires either $e^{ik_{x3}L}$=1 and $e^{ik_{x4}L}$=1 or $e^{ik_{x3}L}$=$-$1 and $e^{ik_{x4}L}$=$-$1. Therefore, we have $k_{x3}L$=$m\pi$ and $k_{x4}L$=$n\pi$, where $m$ and $n$ are both odd or even. The resonate condition indicates large barrier width $L$ is needed to generate resonate tunnelling for $0<\alpha<\alpha_c$. Hence, we only observe resonate tunneling within this angle region for $L$=50 in Fig. \ref{fig3}(c).
For $\alpha$$>$$\alpha_c$, the wavevector $k_{x3}$ becomes imaginary while $k_{x4}$ is still real. By numerical calculation, we find the resonate condition becomes $e^{ik_{x4}L}$=$\pm$1, i.e., $k_{x4}L$=$n\pi$ where $n$ is a nonzero integer, which is similar to that in conventional semiconductors \cite{Griffiths}. The condition $k_{x4}L$=$n\pi$ means that the wider the potential barrier, the more the resonate peaks, which can be directly observed in Fig. \ref{fig3}(c) as indicated by the blue dashed line.
For electrons incident with wavevector $k_2$, as shown in Figs. \ref{fig3}(b) and \ref{fig3}(d), $T_1$ is greatly enhanced in comparison to the results induced by a stepped barrier. For a short barrier such as $L$=10 nm, $T_1$ and $R_2$ doesn't decrease to zero immediately when $\alpha$ exceeds the critical angle $\alpha_c$ [see  Fig. \ref{fig3}(b)]. Although the states $k_{x3}$ becomes evanescent wave, it can still penetrate a short barrier. For a wider barrier such as $L$=50 nm, $T_1$ only shows some resonate peaks for $0<\alpha<\alpha_c$. The resonate condition is also $k_{x3}L$=$m\pi$ and $k_{x4}L$=$n\pi$, where $m$ and $n$ are both odd or even.

In order to understand the quantum transport properties of the junctions shown in Figs. 1(a) and 1(b). The ballistic conductances as a function of incident energy for the step and rectangle barriers are plotted in Fig. 4(a)-(d) with various potentials such as $V$=0, 0.1, 0.5, and 0.9, where $G_1$ ($G_2$) denotes the conductance of electrons incident with $-k_1$ ($k_2$), and $G_T$ stands for the total conductance.
As shown in Fig. 4(a), the conductance $G_1$ tends toward saturation with the decreasing incident energy for all barrier cases. The saturation conductance of $G_1$ diminishes as the barrier potential $V$ increases. This decrease in saturation conductance is attributed to the barrier obstructing the normal and specular transmissions, thereby causing an overall reduction in transmission. Concurrently, $G_1$ undergoes mutations at incident energy $E$=$V-E_T$. The magnitude of these mutations becomes more pronounced with larger values of $V$, as the density of states in the transmitted region also undergoes mutation at $V-E_T$. Conversely, as plotted in Fig. 4(b), $G_2$ initially experiences an ascent followed by a descent as the increasing of the incident energy for all barrier cases. $G_2$ takes its maximum around $E $= ($V - E_T$)/2, and gradually diminishes to zero at $E$= ($V - E_T$), giving rise to a negative differential conductance effect. This phenomenon can be elucidated by the fact that, when $E$ is below $V-E_T$, the $k_{x3}$ state in the transmission region becomes an evanescent wave, leading to a sharp decrease in the normal transmission $T_2$ as shown by the blue-dashed line in Fig. 3(e).
Moreover, when $E$ is less than the brim of the Mexican-hat dispersion $E_T$, the incident $k_2$ state becomes an evanescent wave, preventing transmission and resulting $G_2$ going straight to zero.

Now, let's turn back to the total conductance $G_T$ depicted in Fig. 4(c) which is measured experimentally. As discussed in Figs. 4(a) and 4(b), $G_1$ tends to saturate with increasing incident energy, while $G_2$ sharply decreases. Consequently, the behavior of the total conductance $G_T$=$G_1$+$G_2$ versus the incident energy is similar to that of $G_2$. It also exhibits a negative differential conductance (NDC) effect. The transition energy of the NDC effect is $-E_T+V$. Initially, when the potential $V$ is absent, the peak-to-valley ratio of the NDC effect reaches a maximum value of 0.5 (see the black dashed line), but it gradually decreases with increasing $V$. When the electrical potential $V$ exceeds $E_T$, the NDC effect vanishes because there is only normal reflection and transmission in this case. Therefore, it is evident that the NDC effect can be effectively controlled by adjusting the barrier potential $V$.

\begin{figure}[tbp]
\centering
\includegraphics[bb=60 24 549 542, width=0.48\textwidth]{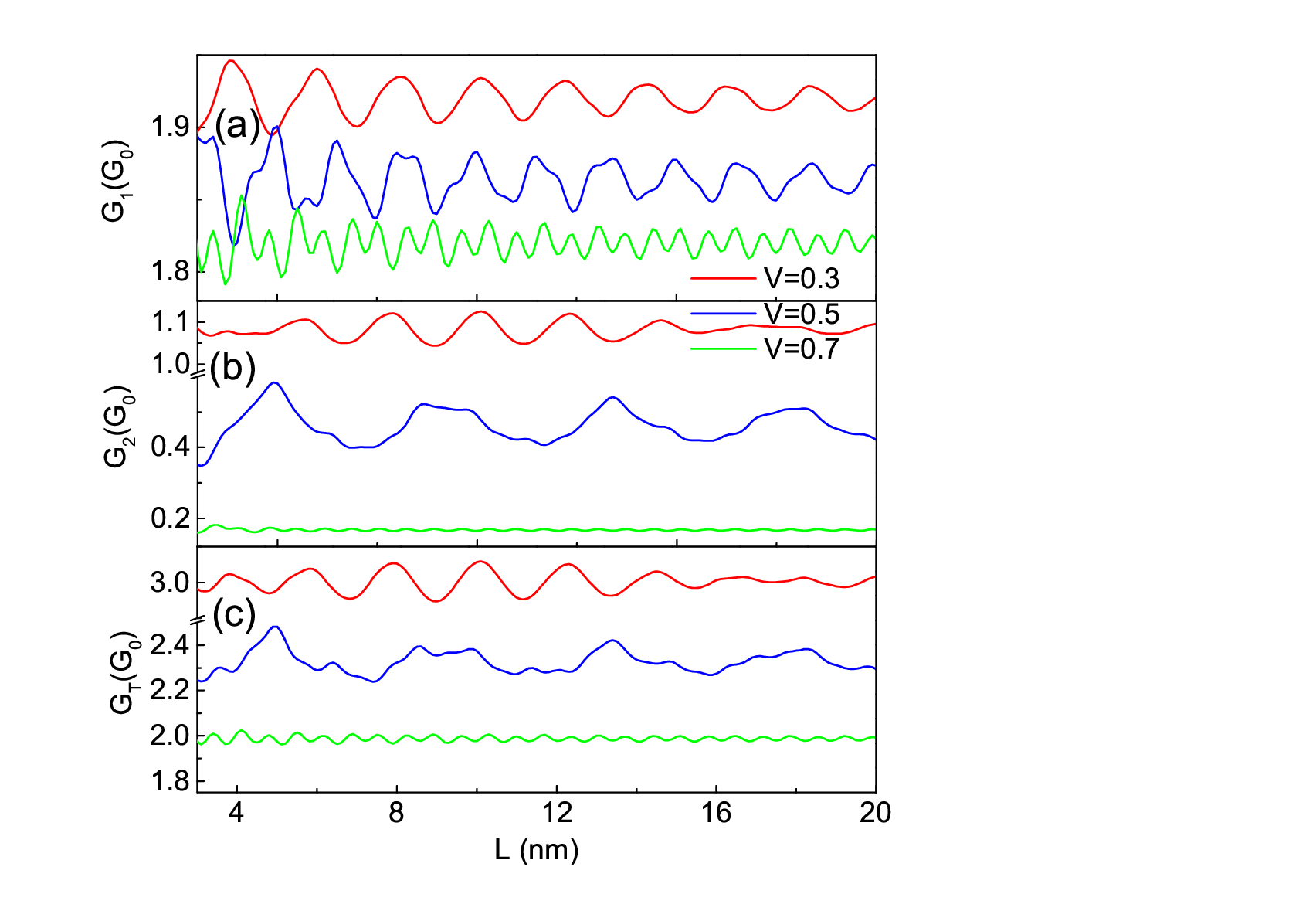}
\caption{(Color online) Conductances as a function of the barrier width for $E$=-0.4 under various potentials, where (a) for $G_1$, (b) for $G_2$, and (c) for $G_T$.}
\label{fig5}
\end{figure}

Figure 5 presents the conductances as a function of the barrier width for $E$=$-$0.4 under various potentials.
As shown in Fig. 5, the conductances can be examined in two distinct scenarios.
In the regime where $V$ $<$ $E - E_T$, both $k_{x3}$ and $k_{x4}$ are real and there are four propagation modes within the barrier region, i.e., two forward and two backward. The oscillations of both $G_1$ and $G_2$ with respect to barrier width $L$ reveal a nuanced dynamic, characterized by the smaller oscillation period of $G_1$ compared to that of $G_2$ as indicated by the red and blue lines in Fig. 5(a) and 5(b). A discernible phase difference further distinguishes these oscillations. As the increasing of the barrier potential $V$, the amplification of this period difference becomes evident, ushering in the emergence of multiple periods in the conductance $G_T$. This phenomenon is particularly pronounced at elevated barrier potentials as shown in the red and blue lines in Fig. 5(c).
As discussed in Fig. (4), the $k_{x3}$ states matches the incident state $k_2$ better compared with the state $k_{x4}$. However, under the condition $V$$>$$E - E_T$, the wavevector $k_{x3}$ becomes imaginary while $k_{x4}$ remains real. Consequently, $G_1$ persists in its oscillations with $L$, while $G_2$ undergoes a substantial diminishment and cessation of oscillation as denoted by the green lines in Figs. 5(a) and 5(b). Consequently, in this distinctive scenario, the conductance $G_T$ is unequivocally dominated by the oscillatory behavior of $G_1$ as indicated by the green lines in Figs. 5(c).

\section{Summary}
In summary, we studied the quantum scattering and transport of
electrons with Mexican-hat dispersion through both step and rectangular
potential barriers by using the transfer matrix method.
Owing to the torus-like iso-energy lines of the Mexican-hat dispersion, we observed the presence of double reflections and double transmissions in both two different barrier scenarios, i.e., the normal reflection (NR), retro-reflection (RR), normal transmission (NT), and specular transmission (ST).
For the step potential with electrons incident from the large wavevector, the transmission is primarily governed by NT with nearly negligible ST, while the reflection is dominant by RR (NR) within (outside) the critical angle. Additionally, for electrons incident from the small wavevector, the NT can be reduced to zero by adjusting the barrier, resulting in a significant enhancement of ST and RR.
For the rectangular barrier, the transmission and reflection spectra resemble those of the step barrier, but there are two kinds of resonant tunneling which can lead to perfect NT or ST. There exists a negative differential conductance (NDC) effect in the conductance spectrum. The conductance and the peak-to-valley ratio of the NDC effect can be effectively controlled by adjusting the height and width of the barrier as well as the incident energy.
Our results provide a deeper understanding of the electron states governed by the Mexican-hat dispersion.

\section{Acknowledgments}
This work was supported by the National Natural Science Foundation of China (Grant Nos. 12374071, 12174100, 12164021 and 11804092), and the Natural Science Foundation of Jiangxi Province (Grant No. 20212ACB201005).


\end{document}